\documentclass{aa}                 
\usepackage{graphicx}
\usepackage{txfonts}

\def\MKA{Mkn~421}
\def\MKB{Mkn~501}
\def\ESA{1ES~1959+650}
\def\ESB{1ES~2344+514}
\def\ESC{1ES~1553+113}
\def\ESD{1ES~1218+304}
\def\PK{PKS~2155$-$304}
\def\PKA{PKS~2005$-$489}
\def\H{H~1426+428}
\def\HA{1H~1100$-$230}

\def\SW{{\it Swift} }
\def\XRT{{\it XRT} }
\def\UVT{{\it UVOT} }
\def\SAX{{\it Beppo}SAX }
\def\XMM{{\it XMM}-Newton }
\def\RXTE{{\it RXTE} }
\def\HESS{H.E.S.S.}
\def\MAGIC{{\it MAGIC}}
\def\nh{$N_{\rm{H}}$}
\def\fluxd{$\times10^{-10}$erg cm$^{-2}$ s$^{-1}$}
\def\fluxu{$\times10^{-11}$erg cm$^{-2}$ s$^{-1}$}
\def\Flux{$10^{-10}$erg cm$^{-2}$s$^{-1}$} 
\begin{document}
\title{SWIFT observations of TeV BL Lac objects}
\author{
A. Tramacere\inst{1} 
\and P. Giommi\inst{2} 
\and E. Massaro\inst{1,2} 
\and M. Perri\inst{2} 
\and R. Nesci\inst{1}
\and S. Colafrancesco\inst{3} 
\and G. Tagliaferri\inst{4} 
\and G. Chincarini\inst{4} 
\and A. Falcone\inst{5}
\and D.N. Burrows\inst{5}
\and P. Roming \inst{5}
\and M. ~McMath Chester\inst{5}
\and N. Gehrels\inst{6} 
}
\institute{
  Dipartimento di Fisica, Universit\`a La Sapienza, Piazzale A. Moro 2, 
  I-00185 Roma, Italy
  \and ASI Science Data Center, ESRIN, via G. Galilei, I-00044 Frascati, Italy
  \and INAF, Osservatorio Astronomico di Roma, Via Frascati 33, Monteporzio Catone, Italy
  \and INAF, Osservatorio Astronomico di Brera, via Bianchi 46, 23807 Merate, Italy
  \and Department of Astronomy and Astrophysics, Pennsylvania State University, USA 
  \and NASA-Goddard Space Flight Center, Greenbelt, Maryland, USA
}

\offprints{~~~~~~~~\\
P.~Giommi: ~paolo.giommi@asdc.asi.it}

\date{Received:.; accepted:.}

\markboth{}
{}

\abstract
{
  We present the results of a set of observations of nine TeV detected BL Lac
  objects performed by the $XRT$ and $UVOT$ detectors on board the $Swift$ 
  satellite between March  and December 2005.
}
{
  We are mainly interested in measuring the spectral parameters, and particularly 
  the intrinsic curvature in the X-ray band.
  }				
{
  We perform X-ray spectral analysis of observed BL Lac TeV objects
  using either a log-parabolic or a simple power-law model .
}				
{
  We found that many of the objects in our sample do show significant  
  spectral curvature, whereas those having the peak of the spectral energies
  distribution at energies lower than $\sim$0.1 keV show power law spectra.
  In these cases, however, the statistics are generally low thus preventing 
  a good estimate of the curvature.
  Simultaneous $UVOT$ observations are important to verify how X-ray spectra
  can be extrapolated at lower frequencies and to search for multiple emission 
  components.
}				
{
 The results of our analysis are useful for the study of possible signatures of statistical
 acceleration processes predicting intrinsically curved spectra and for modelling the SED 
 of BL Lacertae objects up to TeV energies where a corresponding curvature is likely to be present.  
}				


\titlerunning{SWIFT observations of TeV BL Lac objects}
\authorrunning{A. Tramacere et al.}

\maketitle


\section{	Introduction}
All blazars detected at TeV energies are nearby High energy peaked 
BL Lacs (HBL), that is objects with the synchrotron peak in their 
Spectral Energy Distribution (SED) close to or within the X-ray band 
(\cite{Padovani}).
The simultaneous study of these sources at TeV and X-ray energies is 
very important to test the spectral and flux correlations predicted by 
emission models, in particular the Synchrotron Self Compton (SSC) 
scenario.  

The energy distribution of the synchrotron component in the $Log(\nu F_{\nu})$ vs.  
$Log(\nu)$ diagram is generally characterized by a rather smooth broad curvature 
which can be represented by a logarithmic parabola (e.g. Landau et al. 1986, 
Massaro et al. 2004a,b).
In a  recent work on the relations between the parameters of synchrotron
and inverse Compton radiation in the SSC scenario applied to \MKB, (Massaro et 
al. 2006) showed that the intrinsic X-ray spectral curvature results in curved 
TeV spectra.

In  this paper we present the results of a number of \SW  (\cite{Gehrels04}) 
observations of a sample of eight TeV BL Lac objects performed between 
March and December 2005 using the \XRT (\cite{Burrows05}) and $UVOT$ 
(\cite{Roming05}) telescopes.
We found that some of these objects exhibit a remarkable X-ray spectral 
curvature in agreement with previous \SAX results 
(\cite{Massaro04a}, \cite{Massaro04b}, \cite{Costa}, Tagliaferri 
et al. 2003). 
A recent analysis on \XMM~ X-ray data of thirteen HBL sources 
(Perlman et al. 2005) also found evidence for significant intrinsic 
curvature.\\

Simultaneous UV and X-ray observations are important to extend the frequency 
window and to search for the existence of single and multiple non-thermal components.
Under this respect \SW observations provide a unique opportunity for
broad-band studies of this class of non-thermal sources.
The present analysis is of crucial importance to measure the intrinsic X-ray 
spectral curvature, and is devoted to the presentation of new UV-X-ray observations 
of a sample of BL Lac objects  and to their spectral analysis.

\section{\SW~ observations and data reduction}

We present results concerning \SW pointings performed between March and 
December 2005.
The satellite operated with all the instruments in data taking mode but here 
we only consider $XRT$ and $UVOT$ data because our 
sources were not detected by the high energy experiment $BAT$ (\cite{Barthelmy05}).
In addition, we excluded from our analysis all observations for which 
the total number of net counts in the $XRT$ was not sufficient for a reliable 
estimate of the spectral parameters, e.g. exposures shorter than 1000s etc.
The log of X-ray observations is reported in Table 1.     

\subsection{XRT data}
 
All the data were reduced using the $XRTDAS$ software (version v1.8.0) 
developed at the ASI Science Data Center (ASDC) and distributed within the 
HEASOFT 6.0.5 package by the NASA High Energy Astrophysics Archive Research 
Center (HEASARC).

The operational mode of \XRT is automatically controlled by the on-board 
software that uses the appropriate CCD readout mode to reduce or eliminate 
the effects of photon  pile-up.  
When the target count-rate is higher than $\approx$1 cts/s the system is 
normally operated in Windowed Timing (WT) mode whereas the Photon Counting (PC) 
mode is used for fainter sources (for details of the $XRT$ observing modes 
see \cite{Burrows05} and  \cite{Hill04}). 
For our analysis we selected photons with grades in the range 0--12 for PC 
mode and 0--2 for WT mode; we also used default screening parameters to produce 
level 2 cleaned event files.
We found that in several cases some pile-up is nevertheless present in PC mode. 
In these cases we excluded the internal part of the Point Spread Function (PSF) 
in the event selection.
More specifically, spectral data collected in the PC mode were extracted in an 
annular region with an outer radius of 30 pixels while the inner radius was 
chosen according to the prescription of Moretti et al. (2005).
Spectra were binned to ensure a minimum of 20 counts per bin.

A residual feature at $E\approx$0.5 keV, is still present in the best \XRT 
calibration available today (Campana et al. 2006). To avoid artificially 
high $\chi^2$ values and possible biases in spectral parameter estimation, 
in accordance with the \XRT calibration experts, we decided to exclude from 
our analysis the energy channels between 0.4 keV and 0.6 keV (Campana \& 
Cusumano 2006, private communication). 
The corresponding background was estimated in a nearby source-free circular 
region having a radius of 35 pixels.

\begin{table}
\caption{\XRT Observation journal and exposures of TeV blazars.}
\label{tab1}
\begin{tabular}{lccrc}
\hline
 Object  & Date     &  Start UT     & Exp.  & ~\XRT Mode\\
         &          &  h~~m       & ~~~~s~~   &  \\
\hline
 \HA     & Jun 30   &     00:08    & 8521      & PC \\    
         & Jul 13       &    14:19        & 2264      & PC \\
         & Nov 04     &    20:34        & 1164      & PC \\
         & Nov 14     &    14:47        & 554   & PC \\

 \MKA  & Mar 01   & 01:13   & 7134      & WT \\
         & Mar 31   & 00.58   & 6958      & WT \\ 
         & Apr 01   & 04:08   & 2465      & WT \\
         & Apr 29   & 17:45   & 2126      & WT \\
         & May 03   & 05:04   & 318       & WT \\
         & Jul 07   & 19:36   & 2775      & WT \\
         & Oct 04   & 04:19   & 8063      & WT/PC\\
         & Nov 06   & 02:49   & 1126      & WT/PC\\
 
 \ESD    & Oct 30   &    21:18      & 2013      & PC \\    
         & Oct 31   &      19:49      & 3701      & PC \\    
 
 \H      & Mar 31   & 00:02   & 2935      & WT/PC \\
         & Apr 02   & 17:59   & 4646      & WT/PC \\

         & Jun 19   & 04:56   & 21373     & PC \\
         & Jun 25   & 05:54   & 21203     & WT/PC \\
 
 \ESC    & Apr 20   &    03:37        & 5188      & PC\\
         & Oct 06   &      01:16      & 10802     & WT/PC\\
         & Oct 07   &      23:32      & 10718     & WT/PC\\

 \ESA    & Apr 19   & 00:21   & 4437      & WT \\
 
 \PKA    & Mar 31   &   00:06         & 2215      & WT \\    
         & Apr 05   &     00:38       & 8654      & WT/PC \\
         & Apr 06   &      00:45     & 19246     & WT/PC \\
 \PK     & Nov 17   &  19:22          & 1166      & WT/PC \\     
 
 \ESB    & Apr 19   & 00:30   & 4665      & PC \\
         & May 19   & 16:10   & 4039      & PC \\
         & Dec 03   & 00:10   & 12204     & PC \\

\hline

\end{tabular}
\end{table}

\subsection{UVOT data}

A variety of filter combinations and data modes are available for $UVOT$ 
observations.
For fields without bright stars, that is stars bright enough to degrade the 
detector, the most commonly-used  mode observes in six photometric bands: 
$U$, $B$, $V$ and three ultraviolet.  
In some cases our target was quite bright and saturated the image in all 
the photometric bands. 
For this reason not all the optical and UV data could be used for our analysis. 
The list of all available $UVOT$ observations is given in Table 2.

We performed the photometric analysis using a standard tool 
$UVOTSOURCE$ in HEASOFT 6.0.5.
Counts were extracted from a 6$''$ radius aperture in the $V$, $B$,
and $U$ filters and from a 12$''$ radius aperture for the other UV filters
($UVW1$, $UVM2$, and $UVW2$), to properly take into account the wider PSF 
in these bandpasses.
The count rate was corrected for coincidence loss and the background
subtraction was performed by estimating its level in an offset region at 
20$''$ from the source.
The estimate of flux uncertainties is complex due to possible instrumental 
systematics (e.g. residual pile-up in the central region of the PSF) and 
calibration: in particular, the lack of reference stars in the UV bandpasses. 
In this paper we adopt the conservative approach to consider a typical 
uncertainty of 8\% for the $V, B, U$ filters, and of 15\% in the UV band.

The correction for the interstellar reddening was obtained assuming the 
$E(B-V)$ values at the source direction taken from NED and listed in Table 2; 
the fluxes were derived with the same conversion factors given by Giommi et 
al. (2006).

\begin{table}
  \caption{\UVT observation journal of TeV blazars.}
\label{tab1}
\begin{tabular}{lccrc}
\hline
 Object  & Date     &  Start UT     & $E(B-V)$\\
         &          &  h~~m           &  \\
\hline
 \HA     & Jun 30   & 00:06     & 0.059 \\    
         & Jul 13   & 14:19       &  \\
         & Nov 04   & 20:34       &   \\
         & Nov 14   & 14:47        &  \\

 
 \H     & Mar 30  & 23:59        & 0.012 \\
         & Apr 02  & 17:58         &  \\
         & Jun 19   & 06:58       &  \\
         & Jun 25   & 05:58        & \\
 
 \ESC & Apr 20   &   03:37      & 0.052\\
         & Oct 06   &   01:16    & \\
         & Oct 07   &   23:32    & \\

 \ESA    & Apr 19   & 01:03       & 0.177 \\
 
%
 \ESB    & Apr 19   & 01:12      & 0.216 \\
\hline

\end{tabular}
\end{table}

\begin{figure}
{\includegraphics[width=5.9cm,angle=-90]{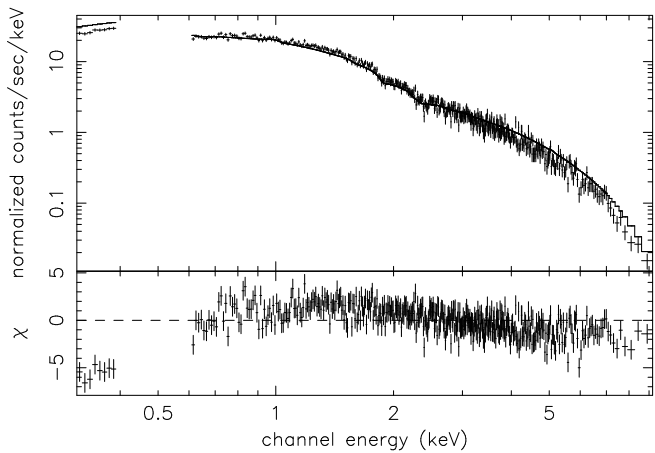}}
\caption{\XRT spectrum of the observation of Mkn~421 performed on 29/04/05.
The systematic deviations of the best fit residuals from a simple power law
and the galactic \nh ~ shows the intrinsic curvature.
}
\label{fig1}
\end{figure}

\section{Spectral analysis and results}

For all the sources the spectral analysis was performed with the \nh~ 
absorbing column densities fixed to the Galactic values and using two 
models for the photon flux: 

 a simple power law (PL)
\begin{equation}
  F(E) = K ~  E^{-\alpha}
\end{equation}

and a log-parabolic law (LP)
\begin{equation}
 F(E) = K ~ E^{-(a + b Log(E))} ~~~~~~,
\end{equation}
where $a$ is the spectral slope (given by the log-derivative) at 1 keV 
and $b$ measures the spectral curvature.
An equivalent functional relationship, useful to obtain independent 
estimates of $b$ and the peak energy  $E_p$ in the Spectral Energy 
Distribution (SED, $S(E)=E^2 F(E)$), is (LPS):
\begin{equation}
S(E)= K_S 10^{-b~[Log(E/E_p)]^2}
\end{equation}
where $K_S=E_p^2 F(E_p)$.

Eq.(2) was used when the residuals with respect to the power law best 
fits showed significant systematic deviations, indicating an intrinsic 
curvature: as an example the spectral data of Mkn 421 are shown in Fig.1.
Typically, such deviations from a simple power law were found in the long
exposures where the total counting statistics were high enough for a good 
description of the spectral shape. 
In these cases the use of the LP law improved significantly the 
$\chi^2$ and eliminated the model systematics of the residuals. 

In principle there are several possible curved models that could give 
acceptable fits to the data. 
For this analysis we apply the LP model for the following reasons: 
$i$) \SAX spectral analysis of several blazars has shown that this
model describes the well data over a range from $0.1$ keV up to 
$\sim$100 keV (\cite{Massaro04a}, \cite{Massaro04b}, \cite{Giommi05} 
\cite{Donato}); 
$ii$) log-parabola is the simplest analytical law to describe a continuous 
curvatures and, in particular, it implies a linear relation between 
the spectral slope and the logarithm of energy;
$iii$) this spectral distribution is based on a physical interpretation 
in terms of statistical acceleration of the emitting particles 
( \cite{Massaro04a},\cite{Massaro06});
$iv$) the curvature parameter allows to compare the spectral evolution 
of the sources in a homogeneous way; this comparison would result more 
complex if the curvature is described by two or more parameters.

An alternative possibility that would probably give technically acceptable 
fits in the limited \XRT energy range is to allow \nh~ to be a free parameter . 

 We verified that in some sources a PL best fit with a free absorbing 
 column yields \nh~ values well in excess of the galactic ones and 
 significantly variable over time scales as short as one day.
 Such a variation is,however, unlikely for absorbing columns distributed over 
 large distances.
 Moreover a large intrinsic absorption in the X-rays would imply a strong 
 extinction in the UV band in contrast with the \UVT data reported below.

 A summary of the results of this spectral analysis is reported in Table 3, 
 for the sources for which we applied the LP model (or LPS for a direct 
 estimate of $E_p$), and in Table 4 for the sources for which the PL gave 
 acceptable fits.  
 For the few observations in which very low best fit curvatures were obtained
 the spectral parameters are given in both Tables 3 and 4.

In the following we present our results for each source.

\subsection{\HA}
At $z$=0.186,  \HA~is one  of the farthest  TeV BL Lac  objects. 
This source was observed by \SAX twice in 1997 and 1998 and showed a 
stable curvature $b=0.3\pm0.1$, although its flux, SED peak  position, 
and spectral slope at 1 keV $a$ were found to be variable.  
In the two states the 2--10 keV flux changed from 2.5 to 3.7\fluxu, 
$a$ moved from $1.9\pm0.1$ to $1.6\pm0.1$ and $E_p$ from 1.5  
to 4.6 keV, respectively (\cite{Giommi05}).   
The source was observed by H.E.S.S. in 2004 and 2005. 
(\cite{Aharonian06a}) and  its TeV  spectrum, after the  correction for
pair  production absorption,  showed  an  intrinsic photon  index
around the rather flat  value of  1.5 when using  a low  intensity EBL
model.  
Our analysis of  \SW data shows a significant  spectral curvature only
during  the  longest   observation  on  June  30,  with   a  value  of
$b=0.40\pm0.06$, which is statistically  consistent with those found by
{\it Beppo}SAX and \XMM~ (Perlman et al. 2005).
The much shorter observation on July 13 is compatible
with a  low curvature spectrum,  and the PL  fit gave a  good $\chi^2$
(see  Table  4).  
In  the  other two  pointings  of  November 2005  the
statistics were too low to measure any spectral curvature.
 
The 2--10  keV flux  was remarkably stable  during all  the pointings,
around the value of 4\fluxu.  
The source was detected by \UVT and the
optical-UV photometry on June 30 and  July 13 gives the
flux values reported in Table 5. 
Fig. 2 shows the SED of \HA~ from the IR to X-ray band, including 
literature measurements and \SAX data from Giommi et  al. (2005).  
Note  the very good agreement  between optical and UV data at different 
epochs  while X-ray data show a much stronger variability.  
We found that the low energy extrapolation of the LP law 
derived only from X-rays generally fails to match UV points, and therefore
we fitted simultaneously \UVT and \XRT data.
The new values of the spectral parameters were different 
from those obtained from X-ray data only.  We verified that these new values
were statistically consistent with the \XRT data by computing again the $\chi^2$ 
keeping $a$ and $b$ frozen to the new values.
These results are also given in Table 3 with the label UV.

\begin{figure}
\resizebox{\hsize}{!}
{\includegraphics[height=8cm,width=8cm,angle=0]{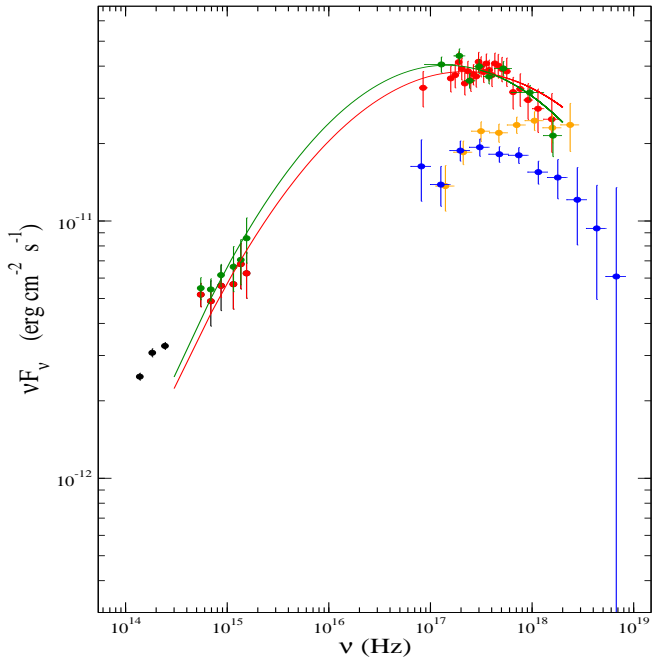}}
\caption{Spectral Energy Distributions of \HA~ derived from \XRT and \UVT 
observations on 30 June 2005 (filled circles) and on 13 July 2005 (filled squares).
For comparison we plotted also \SAX data (triangles) and IR photometric 
points from 2MASS (open circles). 
Dashed lines correspond to simultaneous \XRT UV-\UVT fit.
X-ray open triangles correspond to the \SAX observations performed on 4 January 1997 
and 19 June 1998).
}
\label{fig2}
\end{figure}

\subsection{\MKA}
\MKA, one  of the nearest  BL Lac objects  at $z$=0.031, was  the first
extragalactic  source  to be detected at  TeV  energies (\cite{Punch}).  
It shows strong variability  both at X-ray and TeV energies, with time
scales ranging from less than one hour to years (e.g. Gaidos et al. 1996).   
Spectral variability was observed in the TeV band by \cite{Kren} using 
the Whipple  telescope in  2000/2001 who also discovered a correlation 
between flux and spectral index when averaging the observations over the 
whole data set.  
In the X-ray band  \MKA~ shows strong evidence for intrinsic spectral
curvature that can be described very well by a LP law (e.g. \cite{Massaro04a}).
 
\MKA~ was  observed by  \SW eight times  from March to  November 2005,
however, the last  two observations were performed mainly  in PC mode,
and pile-up was too severe for  a reliable analysis.  
Large variations of the  2--10 keV flux,  by a factor  of about 20, were
observed from March to July when it reached  the highest level of $6.7$\fluxd; 
in the two subsequent  observations it  decreased to $\sim$4.5\fluxd.   
In all the observations the spectrum was remarkably curved with $b$ ranging
from  about  $0.3\pm0.03$  to  $0.46\pm0.02$; the peak
energy in the SED changed between 0.1 and 0.8 keV.
Three SEDs of \MKA~ are shown in Fig.3, unfortunately \UVT images are
saturated and no information on the UV variability related to that
observed at X-ray is available.

 There is a wide literature on past X-ray observations of this source
 (e.g. Massaro et al. 2004a, Tanihata et al. 2004 and references therein).
 In our \XRT observations \MKA~ showed large flux variations that are 
 consistent with historic data. 
 The highest value of flux in the 0.2--10.0 keV band recorded in July
 2005 is compatible with those reached on April 2000 (Massaro et al. 2004a), 
 but the curvature is higher with a value of about $0.37$ while in 
 April 2000 it was about $0.2$. 
 Typical $E_p$ values are different in July 2005 from those of April 2000 
 $E_p$ was about $3$ keV.
 The description of the dynamic of the X-ray spectral evolution is a complex
 topic and an exhaustive analysis will be presented in a future paper 
 (Tramacere et al. 2006, in preparation).

\begin{figure}
\resizebox{\hsize}{!}
{\includegraphics[height=8cm,width=9cm,angle=0]{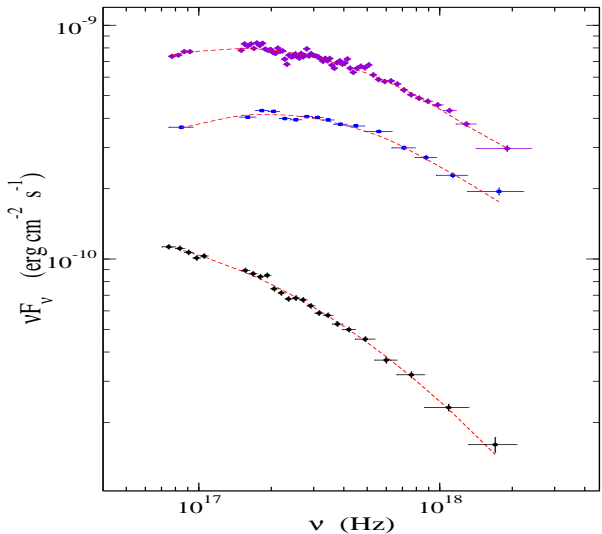}}
\caption{
Spectral Energy Distributions of Mkn~421 in the \XRT energy range 
in three different brightness states observed on March 01 (bottom), 
April 29 (centre), and July 7 2005 (top).
}
\label{fig3}
\end{figure}

\subsection{\ESD}
This source ($z$=0.182) has been detected at TeV energies in January 
2005 by \MAGIC~ (\cite{Albert06}).
The TeV spectrum is described by a power law with a steep photon index 
of about $3$, very similar to that found for \PKA~ by \HESS~
(\cite{Aharonian05a}).
\SAX observed \ESD~ on 12 July 1999 and found a curvature $b=0.37\pm0.03$ 
and a flux about 1.5\fluxu (\cite{Giommi05}).

\SW observed this object in October 2005 and found a flux level similar to 
that observed by \SAX but the statistics were too low to measure the curvature 
with a sufficiently small uncertainty.
PL best fit indicates a relatively high variation of photon index 
corresponding to a flux decrease of about 50\%.

\subsection{\H}
\H~  is one  of  the most  extreme  HBL objects:  it  was observed  in
February  1999  by   \SAX  and  showed  a  flat   power  law  spectrum
($\alpha_X$=0.92$\pm$0.04)  which  placed   the  synchrotron  peak  at
energies higher  than $\sim$100 keV (\cite{Costa}) similar to \MKB~
and \ESB~ in high states.  
At  TeV energies it was observed  by the CAT telescope during 1998-2000
(\cite{Djannati}) and no significant evidence of  spectral variability has
been found (\cite{Aharonian02}).  
This source was observed by \XMM in 2001 and Blustin et al. (2004) 
investigated the presence of intrinsic absorbers using the High Resolution
Reflecting Grating Spectrometer (RGS) and did not find evidence for
broad absorption features nor for narrow emission and absorption lines.  
It was also observed in 2002 by \RXTE in the 3--24 keV  band and showed 
a significant variability  both in flux and spectra (\cite{Falcone}). 
Sometimes these authors observed a spectral index implying  a 
synchrotron peak in excess  of $\simeq$ 100 keV while at other times  
they observed the peak in the  2.9--24 keV band.
 
\SW observed  this source four times  (see Tab. 1).  We  fixed \nh~ to
the Galactic  value of $1.38\times10^{20}$cm$^{-2}$,  however, leaving
it  as a  free parameter  the  value returned  by the  fit was  always
consistent  with  that given above.   
We found evidence for an intrinsic curvature in two cases
out of four:  on  June  19  the source  showed  a  curvature  parameter
$b=0.34\pm0.03$  with $E_p=2.0\pm0.1$  keV,  and on  June  25 $b$  was
$0.37\pm0.05$ and $E_p$ moved to $1.4\pm0.05$ keV. 
The observation on 31 March 2005 is instead compatible with a PL
spectrum with a photon index $\alpha$=$1.96\pm0.06$; the $F$-test
probability for the LP gave a value  of about 0.4 and doesn't help us 
in choosing between  the two spectral  models.  
In this observation the statistics were not rich enough to find 
the actual position of the peak and one cannot exclude that the 
SED could rise above $\sim$10 keV, but the results of our fit are
also consistent with  a spectrum  with a  low curvature  of about
0.1--0.2 and the peak around 1--2 keV.  
Moreover, we also have to take into account that systematic effects due  
the instrument's effective area below 2 keV, combined with the effects 
of the \nh~ absorption and no firm conclusions can be reached.

The observation on April 2 is similar to that of March 31 but with a
better statistics.
In this case the  source was observed  both in PC
and WT mode, but  due to the poor number of counts  of the PC mode, we
discuss here the  results for the fit of the summed  signal of the two
modes.
We estimated a curvature of  $0.12\pm 0.06$ for the LP fit and
a photon index  of  2.05$\pm$0.02 for  PL  fit: the low value of the
$F$-test  probability (see  Tab.4.), however, suggests that the
curved model  could give  a better description  of the data.
In this case the peak position derived by a LPS model was $0.8\pm0.3$ keV.
 


A better determination of the SED is obtained when \UVT data are taken
into account (see Fig.4).   
Note that the optical photometric points ($V$ and $B$) are dominated 
by the flux from the host galaxy and that the non-thermal UV flux 
of \H~ remained substantially stable at variance with the X-ray flux. 
Therefore, a simultaneous fit  in these two  bands is not straightforward.   

In the June 19 observation we found that LP model can give 
a good description  of  the  data  from  UV  to the  X-rays.
We performed a fit on both \UVT and \XRT data with a LP (solid line in 
Fig. 4) and obtained spectral parameters significantly different from those 
obtained fitting only \XRT data (dashed line): 
the new values are reported in Table 3 with the label UV. 
On  April 02 only a single LP can match approximately both X-rays and  
UV points: 
in this case we obtained $b=0.17$, consistent with that found for only \XRT 
data ($0.12\pm0.06$) within the statistical errors.  
The  $\chi^2$ of simultaneous \UVT-\XRT fits are also very satisfactory for 
both observations, confirming  that the addition of  UV data does  
not produce  inconsistencies.  
Note,  however, that the  lowest energy \XRT point in the June 19 observation 
lies significantly below the LP spectrum and is nearly coincident with that 
of the other pointing.  
We do  not have a simple  explanation for this discrepancy: it could simply be  
a  statistical  fluctuation,  but also an indication of the emerging of another 
component at energies higher than $\sim$1 keV.

\begin{figure}
\resizebox{\hsize}{!}
{\includegraphics[height=8cm,width=8cm,angle=0]{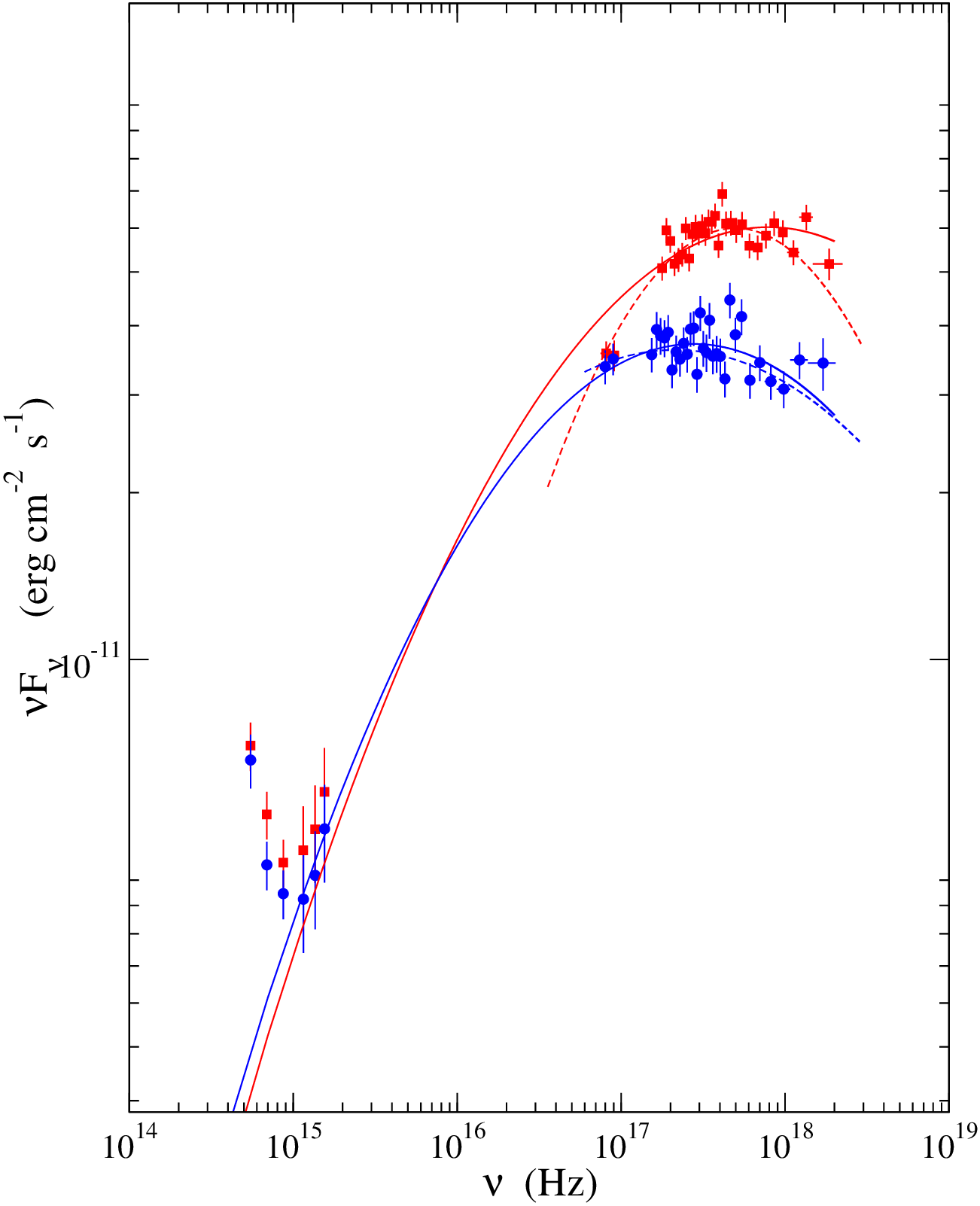}}
\caption{
Spectral Energy Distributions of H 1426+428 in the \XRT \UVT energy 
range in two different brightness states observed on  19 June (squares) and 
02 April (circles) 2005. 
Dashed lines correspond to \XRT data fit. 
Solid lines correspond to simultaneous \XRT \UVT (UV bands only) data fit.
}
\label{fig4}
\end{figure}

\begin{figure}
\resizebox{\hsize}{!}
{\includegraphics[height=8cm,width=8cm,angle=0]{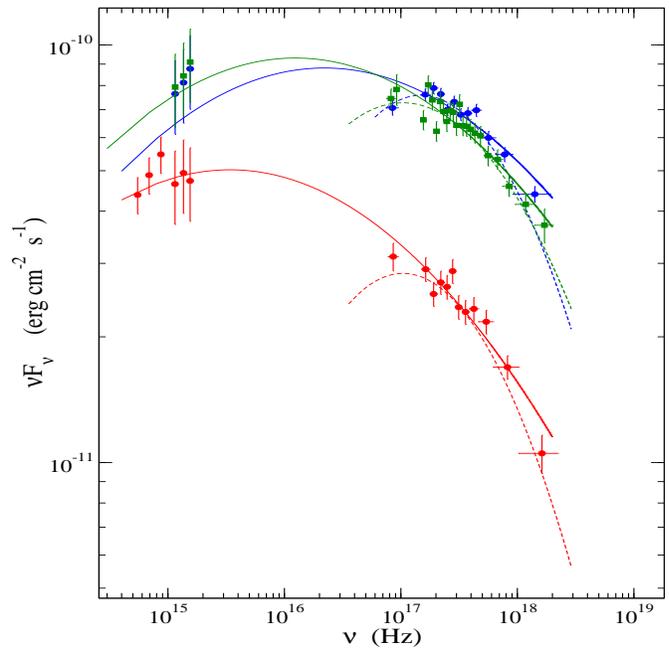}}
\caption{
Spectral Energy Distributions of 1ES 1553+113 observed  on 20/04/05 
(filled circles), 06/10/05 (open diamonds) and 08/10/05(filled squares), 
simultaneously by \XRT and \UVT (UV bands only). 
Dashed lines correspond to \XRT data fit. 
Solid lines corresponds to simultaneous \XRT and \UVT (UV bands only) data fit.
}
\label{fig5}
\end{figure}

\begin{table*}
\caption{Spectral parameters and fluxes of the log-parabolic model  for the 
 \XRT observations of TeV sources}
\label{tab2}
\begin{tabular}{lccccccc}
\hline  
Source/Date &$N_H$                &$a$  &$b$  &$K$                &$\chi^2_r$/dof  &$E_p^*$   &Flux(2-10 keV)  \\ 
            &10$^{20}$cm$^{-2}$   &$ $  &     &$10^{-2}$          &                &(keV)       &\Flux\\ 
\hline

\HA & & & & & & &  \\ 
Jun 30  &5.7  &1.95(0.03)  &0.40 (0.06)	&2.38(0.04)  &0.95/121 &1.15(0.09)  &0.44\\ 
~UV     &     &2.01      &0.15     	&2.30(0.03)  &1.13/123 &---         &--- \\
Jul 13  &     &2.16(0.06)  &0.1  (0.1)	&2.46(0.07)  &0.79/70  &0.3 (0.4)   &0.42\\
~UV     &     &2.08       &0.17     	&2.40(0.06)  &0.82/72  &---         &    \\

\hline

\MKA & & & & & & & \\
Mar 01 &1.61 &2.58 (0.01) &0.31(0.03)  &4.35 (0.03) &1.06/299 &0.12(0.02)  &0.37\\
Mar 31 &     &2.410(0.008) &0.35(0.02)  &7.09 (0.04) &1.097377 &0.26(0.02)  &0.73\\
Apr 01 &     &2.201(0.009) &0.46(0.02)  &18.7 (0.1) &0.97/396 &0.60(0.02)  &2.33\\
Apr 29 &     &2.095(0.008) &0.44(0.02)  &25.6 (0.1) &0.97/442 &0.78(0.02)  &3.76\\
May 03 &     &2.14 (0.03)  &0.41(0.06)  &15.8 (0.3)  &1.16/169 &0.67(0.08)  &2.24\\
Jul 07 &     &2.165(0.005) &0.37(0.01)  &47.61(0.01) &1.22/450 &0.60(0.02)  &6.74\\
\hline

\ESD & & & & & & & \\
Oct 30 &1.73 &1.8 (0.1)	   &0.6 (0.2) &0.17(0.05)  &1.08/32 &1.5 (0.2)   &0.23\\
Oct 31 &     &2.14(0.06)   &0.3 (0.1) &1.12(0.04)  &1.09/61 &0.6 (0.2)   &0.17\\
\hline

\H & & & & & & & \\
Mar 31 &1.38 &1.96(0.06)   &0.16(0.1)   &0.67(0.03)  &1.07/70   &1.3(0.5)    &0.16\\ 
Apr 02 &     &2.01(0.03)   &0.12(0.06)  &1.48(0.02)  &0.95/215  &0.8(0.3)    &0.33\\ 
~~UV     &     &1.9         &0.17       &1.47(0.02)  &0.96/217  &---         &    \\ 
Jun 19 &     &1.78(0.02)   &0.34(0.03)  &2.27(0.02)  &0.92/300  &2.0(0.1)    &0.56\\
~~UV     &     &1.84       &0.16      &2.20(0.02)  &1.03/302  &---         &    \\
Jun 25 &     &1.89(0.02)   &0.37(0.05)  &1.85(0.02) &1.15/232  &1.4(0.05)   &0.38\\
\hline

\ESC & & & & & & & \\
Apr 20 &3.67 &2.25(0.05)   &0.34 (0.09) &1.60(0.04)  &1.14/100	 &0.4 (0.1)   &0.21\\
~~UV   &     &2.3         &0.08       &1.54(0.03)  &1.21/102	 &---         &    \\
Oct 0 6WT&   &2.20(0.02)   &0.32 (0.05) &4.84(0.07)  &0.98/206	 &0.50(0.09)  &0.69\\
~~UV   &     &2.16        &0.08       &4.36(0.04)  &1.11/208	 &---         &    \\
Oct 06 PC&   &2.14(0.02)   &0.34 (0.05) &4.57(0.06)  &0.93/229	 &0.61(0.08)  &0.69\\
Oct 08 &     &2.17(0.02)   &0.23 (0.05) &4.22(0.06)  &0.89/205	 &0.4 (0.1)   &0.67\\
~~UV   &     &2.21        &0.08      &4.12(0.05)  &0.93/205	 &---         &    \\
\hline

\ESA  & & & & & & & \\
Apr 19 &10.0 &2.09 (0.02)  &0.46(0.03)  &8.15 (0.06) &0.98/282 &0.79(0.05)  &1.17\\  
~~UV     &     &2.21     &0.22      &8.21 (0.05) &1.21/284 &----        &    \\  
\hline

\PKA & & & & & & & \\
Mar 31  &5.08 &2.90(0.03)   &0.27 (0.09) &1.85(0.04)  &0.94/126 &---           &0.11\\
\hline 

\PK & & & & & & & \\
Nov 17  &1.69 &2.70(0.05)   &0.3 (0.1)   &2.26(0.08)  &0.62/53  &0.1(0.3)    &0.16\\
\hline 

\end{tabular}

(*) estimated from the LPS spectral model

\end{table*}

\begin{table*}
\caption{Spectral parameters and fluxes of the power-law model for the 
\XRT observations of TeV sources}
\label{tab3}
\begin{tabular}{lcccccc}
\hline
Source/Date           &$N_H$    &$\alpha$  &$K$    &$\chi^2_r$/dof &P($F$-test) & Flux(2-10 keV) \\
                      &10$^{20}$cm$^{-2}$   &      & $10^{-2}$  &   PL  &    &\Flux \\
\hline

\HA & & & & & & \\
Jul 13  &5.7   &2.03(0.04)  &2.46(0.07)  &0.81/71   &0.187 &0.44\\
Nov 04  &      &2.15(0.07)  &2.0(0.1)    &1.10/25   &      &0.42 \\
Nov 14  &      &2.1(0.1)    &1.6(0.1)    &1.13/15   &      &0.41 \\
\hline

\ESD & & & & & & \\
Oct 30  &1.73  &2.00(0.06)  &1.11(0.05)  &1.25/33   &0.032 &0.19 \\  
Oct 31  &      &2.23(0.04)  &1.06(0.03)  &1.20/62   &0.016 &0.29 \\
\hline

\H  & & & & & & \\
Mar 31  &1.38  &2.01(0.05)  &0.66(0.02)  &1.08/71   &0.421 &0.17 \\
Apr 02  &      &2.05(0.02)  &1.45(0.03)  &0.97/216  &0.03  &0.35 \\
\hline 

\PKA  & & & & & & \\
Apr 05  &5.08  &3.16(0.02)  &1.19(0.04)  &0.84/200  &0.06 &0.06 \\
Apr 06  &      &3.05(0.02)  &0.99(0.02)  &0.82/217  &0.06 &0.06 \\  
\hline

\ESB & & & & & & \\
Apr 19  &16.3  &2.35(0.06)  &0.74(0.03)  &0.68/48   & &0.11 \\
May 19  &      &2.18(0.08)  &0.67(0.05)  &0.73/27   & &0.13 \\
Dec 12  &      &2.32(0.05)  &0.60(0.25)  &0.83/65   & &0.09 \\
\hline

\end{tabular}  
\end{table*}

\subsection{\ESC}
This object was detected for the first time at TeV energies in 2005 by \HESS~ 
(\cite{Aharonian06b}). 
Its redshift is unknown but recently Sbarufatti et al. (2006) estimated a lower
limit $z>$0.09
\ESC~ was observed by \SAX on 05 February 1998, and showed a strong
spectral curvature with $b$=$0.6\pm0.1$ and a 2--10 keV flux of $1.3$\fluxu.

We report here the results of three \SW observations performed in April and 
October 2005.
In every pointing the curvature is well determined: 
on April 20 $b$ was equal to $0.34\pm0.09$ and a similar value was found 
on October 06 (see Tab. 3), whereas the 2--10 keV flux (6.8 \fluxu) was 
more than 3 times higher than that recorded at the first epoch.
Two days later, on October 07-08, the spectral curvature was slightly lower 
while the flux was practically unchanged.

It is very interesting to note that \ESC~ is the only source in our sample 
with a SED having the optical/UV emission higher than X-rays, as shown in 
Fig.5, and this feature was observed in all the pointings.
We plotted two best fit spectra: one for  the X-ray data  only (dashed lines) 
and the other for both \UVT and \XRT points (solid lines).
It is evident that when the UV points are included in the fit the resulting
values of $b$ and $E_p$ are significantly lower: 
the new values of the spectral parameters are also given in Table 3 and are 
denoted by UV, whereas the fluxes in the \UVT filters are listed in Table 5. 
We verified that the new $\chi^2$ values of \XRT data computed with $a$ and $b$  
frozen to the values coming from simultaneous UV-X points remain acceptable 
(see Tab. 3).
Note in particular that the spectral curvature when UV data are 
taken into account is around 0.1, a value consistent with that found by 
Perlman et al. (2005) in high statistics \XMM~ observations.

\subsection{\ESA}
\ESA~ has a redshift $z$=0.047 and was detected at TeV energies in 1999 
(\cite{Nishiyama99}) and confirmed by Whipple (\cite{Holder}) and HEGRA 
(\cite{Horns}).  
It was observed twice by \SAX~ in September 2001 when it showed a significant
spectral curvature (\cite{Taglia}). 
These authors used a LP law and found a value of $b$ of about 0.4.    
In May 2002 \ESA~ was the subject of a multiwavelength campaign in which a
strong TeV/X-ray correlation was found with the remarkable exception of 
the TeV flare on 4 June 2002 which was not associated to any X-ray activity
(\cite{Kraw}).  

We have only one \SW observation of this source.
In our spectral analysis we adopted the Galactic \nh~ value
of  $1.0\times10^{21}$cm$^{-2}$, although the possibility of a high
intrinsic absorption cannot be excluded.
We confirmed the spectral analysis with $b=0.46\pm0.03$, a flux of about
$1.16$\fluxd and a peak energy at $\simeq0.8$ keV.

Similar curvatures were also found by Perlman et al. (2005).

Recently Gutierrez et al. (2006) reported a time resolved spectral analysis of  
\RXTE observations in the 4--15 keV range and found a curvature parameter in
the interval 0.1--0.4 over a period of about four years.

As in the case of \H~ and \ESC~ we found that simultaneous \UVT-\XRT data are  
compatible with a LP law.
In Fig.6 (solid line), we plotted the LP broad band best fit, which peaks 
at $\sim$0.3 keV and has $b\simeq$0.2.
The $\chi^2$ for only the \XRT data is increased but still acceptable, as for the 
previous sources.

\begin{figure}
\resizebox{\hsize}{!}
{\includegraphics[height=8cm,width=8cm,angle=0]{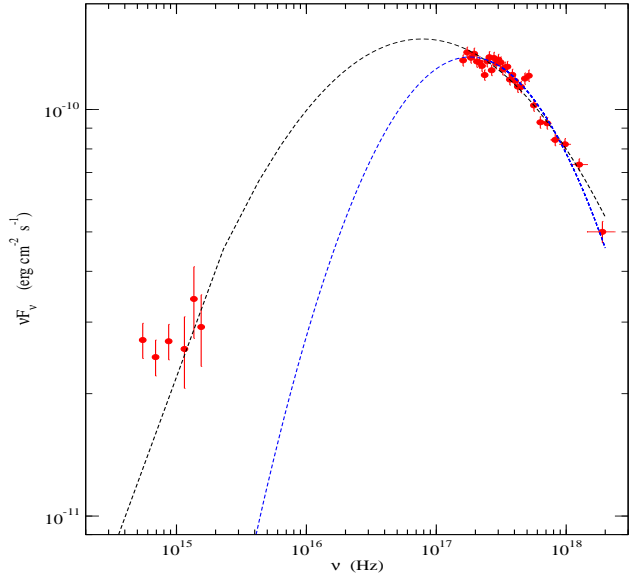}}
\caption{
Optical to X-ray SED of \ESA~ as observed on 19 April 2005 by \SW. 
Dashed line corresponds to \XRT data fit. 
Solid line corresponds to simultaneous \XRT - \UVT (UV bands only) 
data fit.
}
\label{fig6}
\end{figure}

\begin{figure}
\resizebox{\hsize}{!}
{\includegraphics[height=8cm,width=8cm,angle=0]{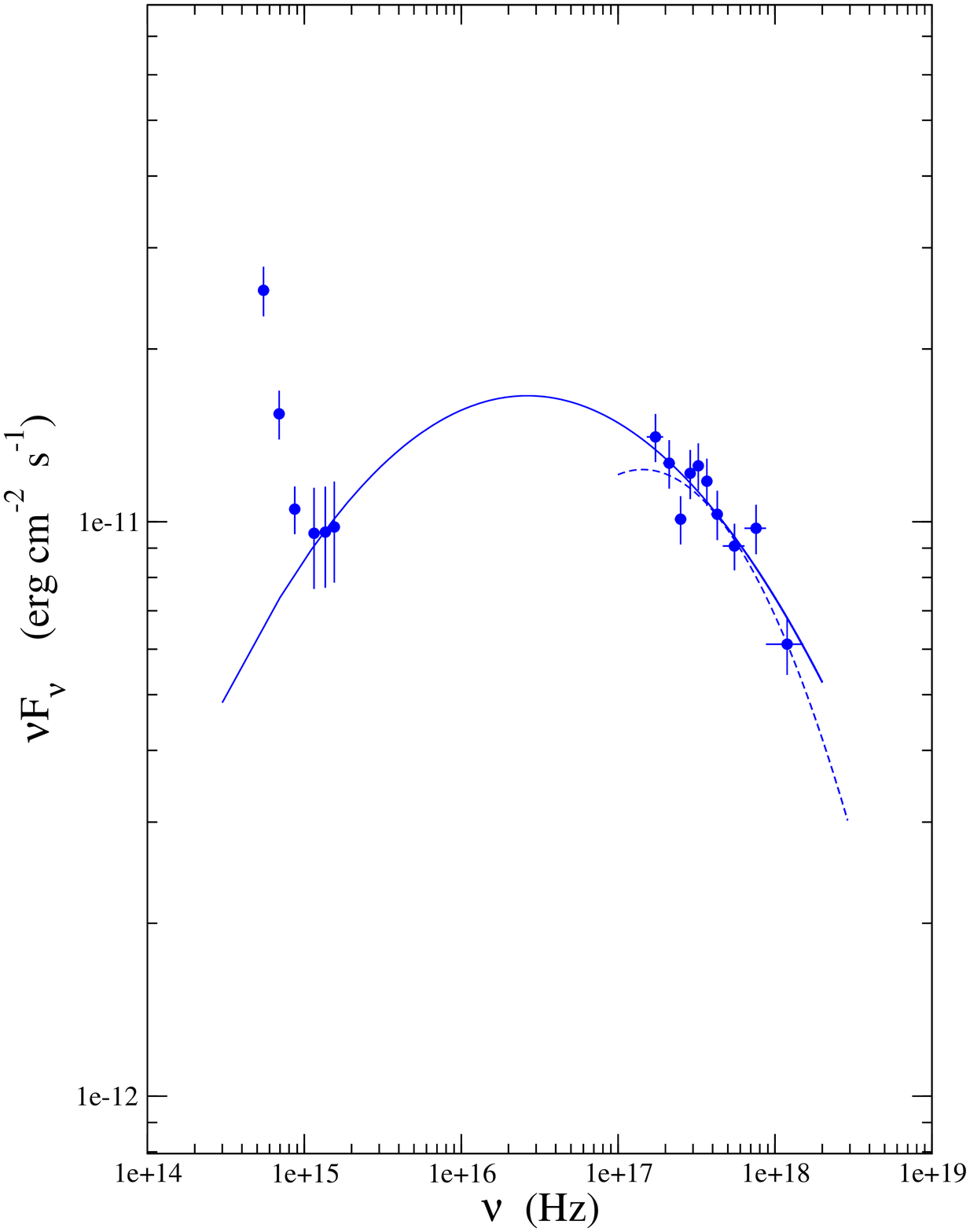}}
\caption{
Optical to X-ray SED of \ESB~ as observed on 19 April 2005 by \SW.
Dashed line corresponds to \XRT data fit. Solid line corresponds to simultaneous 
\XRT \UVT (UV bands only) data fit.}
\label{fig7}
\end{figure}

\begin{table*}
  \caption{Dereddened fluxes (in mJy) in the \UVT bandpasses.}
\label{tab1}
\begin{tabular}{lcrrrrrr}
\hline
 Object  & Date     &  $F_V$    & $F_B$    & $F_U$    & $F_{UVW1}$ & $F_{UVM2}$ &
$F_{UVW2}$ \\
         &          &           &          &          &            &            &         
  \\
\hline 
 \HA     & Jun 30   &  0.94     &  0.71    &  0.64    &   0.49     &   0.50     &   0.40  
 \\    
         & Jul 13   &           &          &          &            &            &         
  \\
 \H      & Apr 02   &  0.79     &  0.40    &  0.29    &  0.21      &   0.20     &   0.21  
  \\
         & Jun 19   &  0.84     &  0.50    &  0.32    &  0.26      &   0.24     &   0.24 
\\
 \ESC    & Apr 20   &  8.0      &  7.1     &  6.3     &  4.0       &  3.6       &  3.1  \\
         & Oct 06   &           &          &          &  6.6       &  6.0       &  5.7  
\\
         & Oct 08   &           &          &          &  6.9       &  6.3       &  5.9   
\\
 \ESA    & Apr 19   &  4.9      &  3.6     &  3.1     &  2.2       &  2.5       &  1.9    
\\
 \ESB    & Apr 19   &  4.6      &  2.2     &  1.2     &  0.8       &  0.7       &  0.6    
\\
 
%
\hline

\end{tabular}
\end{table*}

\subsection{\PKA}
\PKA~ ($z=$0.071) was discovered at TeV energies by \HESS~ in 2004 
(\cite{Aharonian05a}) with an integrated flux over 200 GeV of  
$6.9\times10^{-12}$ erg cm$^{-2}$s$^{-1}$, corresponding to about 25\% 
of the Crab Nebula.
Past X-ray observations were performed by \SAX in September 1996 
and November 1998.  
The curvature measured in those pointings was generally very low equal, 
to $0.05\pm0.11$ and $0.12\pm0.02$, respectively, and the corresponding 
2--10 keV fluxes were $6$ \fluxu and $1.8$ \fluxd (\cite{Giommi02}).

\SW~ observations were performed in March and 
April 2005 and showed again low fluxes and curvatures.
On March 03 we found $b=0.23\pm0.05$ and a flux of about $7$\fluxu,
while in April the spectrum was consistent with a PL with a soft 
photon index (see Table 4).  
The flux computed by this spectral model is about an order of magnitude 
lower than that recorded in April.  
It is interesting to note that such a very soft X-ray spectrum corresponds 
well with that found by \HESS~ (\cite{Aharonian05a}) in the TeV band, for
which a photon index of $\approx$4 was measured, the softest for a BL 
Lac at these energies.  
These authors were confident that it was unlikely to have a high
absorption by EBL and argued that the X-ray spectrum must really be
soft. Our analysis of \SW data supports this conclusion. 

\subsection{\PK}
\PK~ ($z$=0.117) is one of the first BL Lac object detected at TeV energies 
(\cite{Chad99}). 
During October and November 2003 it was observed by \HESS~  and its TeV 
spectrum was found to be compatible with a PL with photon index of 
$3.32\pm0.06$ (\cite{Aharonian05c}). 
\SAX observed this source three times and the spectral curvature was
found in the narrow range $0.27-0.3$, whereas the flux changed 
from $2.5$ to $8.3$ \fluxu (\cite{Giommi02}).

\SW observed \PK~ in 2005 only once and the statistics of 
the data were not sufficient to provide a good estimate of $b$ (see Table 3)
which was found to be compatible with that measured during the previous 
observations, despite the lower flux level.
Note, however, that the value of $a$ was significantly higher than 2
confirming the soft SED of this source.

\subsection{\ESB}
This BL Lac object ($z$=0.044) was detected at TeV energies by Catanese et 
al. (1998), and more recent observations have been presented by Schroedter et al. 
(2005). 
It was observed by \SAX in December 1996 and June 1998 and showed a strong 
X-ray variability with large flux and spectral changes (\cite{Giommi}).

\ESB~ was observed by \SW on three occasions: its 2--10 keV flux was 
quite steady at a typical level of about 1\fluxu.  
The low statistics of these observations and the high value of the Galactic 
column density \nh=$1.63\times10^{21}$cm$^{-2}$ does not allow us to determine 
the value of $b$ with a good accuracy and 
PL best fit parameters are given in Table 4.
Note that the photon index was always larger than 2., indicating
that the peak of the SED is at energies lower than 1 keV, as in the faint
state reported by Giommi et al. (2000).

An \UVT observation is available only for the April 19 pointing:
the stars in the $UVW2$ band image appear elongated, probably from aspect 
reconstruction problems, but still within the 12$''$ photometric radius. 
The results are reported in Table 5.
 Because of the large galactic extinction ($E(B-V)$=0.216), the flux in the M2 
band is substantially affected by the 2200 \AA~ interstellar band.
Assuming the host galaxy estimate from the HST observation (Urry et al. 2000), 
we found 
{\it that it contributes for about 90\% of the signal }
in the $V$ and $B$ filters, and therefore these data are not representative 
of the AGN flux. 
In the UV filters, where the galaxian contribution is much lower, the SED 
appears substantially flat.
In Fig. 7 the dashed line corresponds to LP fit of \XRT data, which is not 
reported in Tab.3 because of the low accuracy of the resulting parameters;
the solid line corresponds to the fit of \UVT and \XRT data with a single LP law,
having a curvature of about 0.15 and $E_p\simeq2.5 keV$.

\section{Discussion}

Our \SW observations of TeV-detected blazars strongly confirm that 
the SED of these sources is often curved and well approximated by a LP
law with peak energy generally in the X-ray band but that can vary 
more than one order of magnitude.

The estimate of the curvature parameter, however, is often a complex 
task as it may be affected by systematic and statistical uncertainties 
that sometimes can be very important, especially when working on a 
rather narrow frequency range.
Under this respect the \SW instrumentation offers the unique possibility 
to simultaneously measure the X-ray and UV fluxes, where the 
contribution from the host galaxy is generally negligible. 
These data can be used to model the SED over a frequency interval 
spanning over three decades allowing us to test whether a single 
emission component can describe all data.
This problem was already apparent from the spectral modelling of the \SAX 
X-ray observations of Mkn 421 (Massaro et al. 2004a) and Mkn 501 (Massaro 
et al. 2004b), from which it resulted that no simple connection between 
optical and X-ray points was possible.
For the sources presented here, we found that a single component is 
generally acceptable. 
However, there are some interesting differences among these sources.

As noted in Sect. 3.5, \ESC~ is the only source having a SED dominated by 
the UV emission: when a single LP is used the peak position moves at 
frequencies lower than 10$^{16}$ Hz. 
UV points show a rather satisfactory agreement with the model whereas
optical data are aligned along a harder spectrum.
Such a discrepancy between optical and UV fluxes, which  could be 
affected by a larger calibration uncertainty, {\it however, it cannot be 
entirely } explained in this way because it is not present in other 
sources and suggests a contribution from a different origin.
Sbarufatti et al. (2005) did not find evidence of a host galaxy
and therefore we must conclude that the $B, U$ excess is
originated in the nuclear environment, although its nature is unclear.

Another very interesting source is \H~ which shows a significantly larger 
variability in the X-ray  band than at UV frequencies. 
Note that the flux at $\sim$0.3 keV is also rather stable
in comparison with that at energies higher than 1 keV, a behaviour 
that recalls that of Mkn 501 (Massaro et al 2004b).
This finding can imply that optical/UV photons are originated from the 
low energy tail of the same electron population that produces X-rays 
photons.  
However, the possibility that the UV radiation is emitted by 
another electron population, rather stable and of lower energy, 
non-cospatial to the high-energy one, cannot be excluded.
To discriminate between these two possibilities it is useful the study 
of its time behavior at different wavelengths. 
We stress that optical and IR observation are not very useful because
of the presence of a bright host galaxy that is very well detectable in 
these bands (see Fig. 4).

Finally, note that all five sources for which the PL model gave an
acceptable fit are characterized by photon indices larger than
2, corresponding to $E_p$ values lower than 1 keV. 
Such steep spectra generally correspond to a low number of net counts in the
high energy channels making curvature very hard to estimate.

The shape of X-ray spectra reflects the energy distribution of the emitting particles. 
The relations between the curvatures of the electron spectrum and those of the 
Synchrotron and Inverse  Compton radiation that they emit has been studied by Massaro et.  (2006).  
Curvature in the electron energy distribution may be the result of statistical acceleration (see 
Massaro et. al 2004a), therefore the X-ray spectral curvature presented in this paper may be used to investigate possible signs of statistical acceleration arising in different conditions like 
energy dependent acceleration probability or 
fluctuations in the energy gain. (Kardashev 1962, Tramacere et al. 2006 in prep.).

An interesting consequence of intrinsic spectral curvature is that it has direct implications 
in the estimate of the pair production opacity due to Extragalactic Background Light (EBL) 
as pointed out by Massaro et al. (2006) for the case of Mkn~501 on the basis of simultaneous 
X-ray and TeV observations. 
The observed spectra are the combination of the intrinsic shape of the IC radiation and 
the $\gamma$-ray opacity. At present the amount of EBL is still uncertain (see Schroedter 2005 for a recent compilation) and upper limits on it, derived assuming a PL spectrum for the TeV emission of nearby BL Lac objects, are reduced if intrinsically curved IC spectra are considered. 

\begin{acknowledgements}

The authors acknowledge financial support by the Italian Space Agency (ASI) to 
the ASDC  and to the Italian Swift team through grant I/R/039/04 and by the 
Phys. Dept. by Universit\`a di Roma La Sapienza. 
This work is sponsored at Penn State University by NASA contract NAS5-00136.
We thank G. Cusumano for his support with the \XRT calibration.
We acknowledge the use of the National Extragalactic Database (NED) for the values 
of  $E(B-V)$ listed in table 2.

\end{acknowledgements}


\begin{thebibliography}{ }

\bibitem[Aharonian et al. 2002]{Aharonian02}
 Aharonian, F., Akhperjanian, A., Barrio, J., et al. 2002,  A\&A, 384, 23

\bibitem[Aharonian et al. 2005a]{Aharonian05a} 
 Aharonian, F., Akhperjanian, A.G., Aye, K.-M. et al.  2005a, A\&A, 436, L17 

\bibitem[Aharonian et al. 2005b]{Aharonian05b} 
 Aharonian, F., Akhperjanian, A.G., Aye, K.-M. et al. 2005b, A\&A, 437, 95

\bibitem[Aharonian et al. 2005c]{Aharonian05c}
 Aharonian, F., Akhperjanian, A.G., Bazer-Bachi, A.R. et al. 2005c, A\&A, 442, 895 

\bibitem[Aharonian et al. 2006a]{Aharonian06a}
 Aharonian, F., Akhperjanian, A.G., Bazer-Bachi, A.R. et al., 2006a,  Nature, 440, 1018

\bibitem[Aharonian et al. 2006b]{Aharonian06b}
 Aharonian, F., Akhperjanian, A.G., Bazer-Bachi, A.R., et al. 2006b,  A\&A, 448,L19

\bibitem[Albert  et al. 2006]{Albert06}
 Albert, J., Aliu, E., Anderhub, H. et al., 2006 ApJ 642, L119 
 
\bibitem[{Barthelmy et al. 2005}]{Barthelmy05}
 Barthelmy, S., Barbier, L.~M., Cummings, J., {et~al.} 2005, SSRv., 120, 95

\bibitem[Blustin et. al  2004]{Blustin}
 Blustin A.J., Page M.J. \& Branduardi-Raymont G. 2004, A\&A, 417, 61

\bibitem[{Burrows et al. 2005}]{Burrows05}
 Burrows, D., Hill, J.~E., Nousek, J.~A., {et~al.} 2005, SSRv., 120, 165

\bibitem[Campana et al 2006]{Campana}
 Campana, S., Beardmore, A.P., Cusumano, G., Godet O. 2006 
  ``Swift-XRT-CALDB-09'' 
  
\bibitem[Catanese et al. 1998]{Catanese} 
 Catanese, M., Akerlof, C.W., Badran, H.M. et al. 1998, ApJ, 501, 616

\bibitem[Chadwick et al. 1999]{Chad99}
 Chadwick, P. M., Lyons, K., McComb, T. J. L., et al. 1999, ApJ, 513, 161 

\bibitem[Costamante et al. 2001]{Costa}
 Costamante, L., Ghisellini, G., Giommi, P. et al. 2001, A\&A, 371, 512

\bibitem[Djannati-Ataj et al. 2002]{Djannati}
  Djannati-Ataj, A. et al., 2002, A\&A, 391, L25 

\bibitem[Donato et al. 2005]{Donato}
 Donato D., Sambruna R. M., Gilozzi M, 2005, A\&A 433 1163 

\bibitem[Draper et al. 2004]{Drap04} 
 Draper P.W., Gray N., Berry D.S. 
  2004, Starlink User Note 214, 15

\bibitem[Dwek \& Krennich 2005]{Dwek} 
 Dwek E., Krennich F.,  2005, ApJ, 618, 657

\bibitem[Falcone et. al 2004]{Falcone}
 Falcone A.D., Cui W. \& Finley J.P. 2004, ApJ, 601, 165 
 
\bibitem[Gehrels et~al. 2004]{Gehrels04}
 Gehrels, N., Chincarini, G., Giommi, P., {et~al.} 2004, ApJ, 611, 1005

\bibitem[Gaidos et~al. 1996]{Gaidos}
 Gaidos J.A, Akerlof C.W, Biller S.D. et al. 1996, Nature, 383, 319 

\bibitem[Giommi et al. 2000]{Giommi}
 Giommi P., Padovani P., Perlman E.S. 2000, MNRAS, 317, 743
  
\bibitem[Giommi et al. 2002]{Giommi02} 
 Giommi, P.,  Capalbi, M., Fiocchi, M. et al,  2002, in "Blazar Astrophysics with {\it BeppoSAX} and Other Observatories", p.  63

\bibitem[Giommi et al. 2005]{Giommi05}
 Giommi P., Piranomonte S., Perri, M., Padovani, P. 2005,  A\&A, 434, 385

\bibitem[Giommi et al. 2006]{Giommi06}
 Giommi P., Blustin A.J., Capalbi M., Colafrancesco S. et al. 
 2006  A\&A, in press 

\bibitem[Gutierrez et al. 2006]{Guti}
 Gutierrez, K., Badran, H.M., Bradbury, S.M. et al. 2006 ApJ, 644 742

\bibitem[Hill et al. 2004]{Hill04} 
 Hill, J.E., Burrows, D.N., Nousek, J.A. et al., 2004, SPIE, 5165, 217
  
\bibitem[Holder 2003]{Holder}
 Holder, J. 2003, Proceedings of the 28th International Cosmic Ray Conference,
 ICRC, 5, 2619


\bibitem[Krawczynski et al. 2004]{Kraw} 
 Krawczynski, H., Hughes, S.B., Horan, D. et al. 2004, ApJ, 601, 151 

\bibitem[Krennrich et al. 2002]{Kren}
 Krennrich, F., Bond, I.H., Bradbury, S.M. et al. 2002, ApJ, 575, L9 
  
\bibitem[Landau 1986]{landau} 
 Landau, R., Golisch, B., Jones, T.J. et al. 1986, ApJ, 308, 78

\bibitem[Massaro et al. 2004a]{Massaro04a} 
 Massaro E., Perri M., Giommi P., Nesci R., 2004a, A\&A, 413, 489 

\bibitem[Massaro et al. 2004b]{Massaro04b} 
 Massaro E., Perri M., Giommi P. et al., 2004b, A\&A, 422, 103 

\bibitem[Massaro et al. 2006]{Massaro06} 
 Massaro E., Tramacere A., Perri M. et al. 2006, A\&A, 448, 861
 
\bibitem[Moretti et al. 2005]{Moretti05}
 Moretti, A., Campana, S., Mineo, T. et al. 2005, Proc  SPIE vol 5898, 360
  
\bibitem[Nishiyama et al. 1999]{Nishiyama99}
 Nishiyama, T. 1999, Proceedings of the 26th International Cosmic Ray Conference,
   ICRC 3, 370.
 
\bibitem[Padovani \& Giommi 1995]{Padovani} 
 Padovani P., Giommi P. 1995, ApJ , 444, 567

\bibitem[Perlman et al. 2005]{Perlman}
 Perlman E.S., Madejski G., Georganopoulos M. et al.,  2005, ApJ 625, 727

\bibitem[Punch et al. 1992]{Punch}
 Punch, M., Akerlof, C.W., Cawley, M.F. et al.,  1992, Nature, 358, 477
  
\bibitem[Roming et~al. 2005]{Roming05}
 Roming, P.W.A., Kennedy, T.E., Mason, K.O. et al, 2005, SSRv.,120, 143

\bibitem[Sbarufatti et al. 2006]{Sbarufatti}
 Sbarufatti et al, 2006, AJ, 132, 1-19

\bibitem[Schroedter et al. 200]{Schroedter05}
 Schroedter, M., Badran, H.M., Buckley, J.H. et al., 2005, ApJ, 634, 947

\bibitem[Schroedter 2005]{Schroedter05a}
 Schroedter, M., 2005 ApJ, 628, 617

\bibitem[Tagliaferri et al. 2003]{Taglia}
 Tagliaferri, G., Ravasio, M., Ghisellini, G. et al.,  2003, A\&A, 412, 711

\bibitem{Tanihata et al. 2004}
 Tanihata, C., Kataoka J., Takahashi T., et al., 2004, Apj, 601, 759

\bibitem[Urry et al. 2000]{Urry00}
 Urry, C. M., Scarpa, R., O'Dowd, M.et al., 2000, ApJ, 532, 816


\end{thebibliography}
\end{document}